\documentclass[aps,prb,10pt,twocolumn,showpacs,amsmath,amssymb,superscriptaddress]{revtex4-1}
\usepackage{graphicx,color}
\usepackage{amssymb,amsmath}
\usepackage[colorlinks=true,citecolor=blue]{hyperref}

\usepackage{soul}
\usepackage{color}
\definecolor{purple}{rgb}{0.7,0.0,1.0}

\begin{document}
\title{Electron and phonon drag in thermoelectric transport\\ through coherent molecular conductors}
\author{Jing-Tao L\"u}
\email{jtlu@hust.edu.cn}
\affiliation{School of Physics and Wuhan National High Magnetic Field Center, Huazhong University of Science and Technology, 430074 Wuhan, China}
\author{Jian-Sheng Wang}
\affiliation{Department of Physics, National University of Singapore, 117551 Singapore, Republic of Singapore}
\author{Per Hedeg{\aa}rd}
\affiliation{Niels Bohr Institute and Nano-Science Center, University of Copenhagen, 2100 Copenhagen {\O}, Denmark}
\author{Mads Brandbyge}
\affiliation{Center for Nanostructured Graphene (CNG), Department of Micro- and Nanotechnology, Technical University of Denmark, DK-2800 Kongens Lyngby, Denmark}
\date{24 January, 2016}
\begin{abstract}
We study thermoelectric transport through a coherent molecular conductor
connected to two electron and two phonon baths using the nonequilibrium
Green's function method. We focus on the mutual drag between electron and
phonon transport as a result of `momentum' transfer, which happens only when
there are at least two phonon degrees of freedom. After deriving expressions
for the linear drag coefficients, obeying the Onsager relation, we further
investigate their effect on nonequilibrium transport. We show that the drag
effect is closely related to two other phenomena: (1) adiabatic charge pumping
through a coherent conductor; (2) the current-induced nonconservative and
effective magnetic forces on phonons.
\end{abstract}
\maketitle

\section{Introduction}
The possibility to engineer electron and phonon transport independently in
nanostructures makes them ideal candidate for thermoelectric applications, the
conversion of heat to electricity, and vice
versa\cite{dresselhaus_new_2007,snyder_complex_2008,majumdar_thermoelectricity_2004,mahan_best_1996,godijn_thermopower_1999,paulsson_thermoelectric_2003,reddy_thermoelectricity_2007,WaWaLu07,baheti_probing_2008,DuDi11,li_colloquium:_2012}.
Thermoelectric transport in quantum wells, wires and dots has been the focus of
intense study in the past decades. Recently, it has become possible to measure the
thermopower of molecular junctions, the extreme minimization of
electronics\cite{paulsson_thermoelectric_2003,reddy_thermoelectricity_2007,baheti_probing_2008,DuDi11}.
Although still in its infancy from application point of view, academically,
thermopower has proven useful as a complementary tool to explore the transport
properties of molecular devices. For example, the sign of the thermopower gives
information about the relative position of the electrode Fermi level within the
HOMO-LUMO gap of the
molecule\cite{paulsson_thermoelectric_2003,Koch04,reddy_thermoelectricity_2007,baheti_probing_2008};
the quantum interference
effect\cite{AndreevThermopower,jacquod_coherent_2010-1,bergfield_thermoelectric_2009,finch_giant_2009},
and many-body
interactions\cite{staring_coulomb-blockade_1993,scheibner_thermopower_2005,boese_thermoelectric_2001,dong02}
also show their signatures in the thermopower.

The interaction of electrons and vibrations within the molecule couples charge 
and phonon heat transport. The signature of this coupling in electrical current
has been used as a spectroscopy tool to  unambiguously identify the molecule.
However, many of the early theoretical work on thermoelectric transport in
molecular conductors treat electron and phonon transport separately, within
the linear regime. Recently, there
are more attempts trying to include the electron-phonon(e-ph) interaction, extend the
analysis to the nonlinear regime\cite{Koch04,Lu07,asai_nonequilibrium_2008,galperin_inelastic_2008,leijnse_nonlinear_2010,nakpathomkun_thermoelectric_2010,hsu_seebeck_2011,Sergueev2011,Ren2012,JiangJW2011joule,liu10,Arrachea14,Sanchez2011,Sanchez2014,Zhou2015,Agarwalla2015,Whitney2013,Yamamoto2015}, and consider multi-terminal transport\cite{entin-wohlman_three-terminal_2010,jiang_thermoelectric_2012,Entin15,Jiang2015,Sanchez2010,Sothmann2012,Whitney2016}.
The e-ph interaction modifies the electronic transmission and consequently the
thermopower. Extending to the
nonlinear regime also helps to make connection with the current-induced heating and heat
transport in molecular devices. 

In this paper, we study the nonequilibrium thermoelectric transport through a
model device, connected to two electron and two phonon baths, including the e-ph
interaction within the device. We use the nonequilibrium Green's function
(NEGF) method to take into account the effect of e-ph interaction within the
lowest order perturbation \cite{LOE1}, assuming the interaction is weak.
Thus, our approach does not apply to molecular junctions that couple weakly
to the electrodes\cite{leijnse_nonlinear_2010,Ren2012,Zhou2015}.
In the linear regime, we derive the thermoelectric transport coefficients
including the e-ph interaction.  We pay special attention to the drag coefficients,
whereby a temperature difference between the phonon baths drives an electrical
current between the electron baths, and vice versa. The drag effect has been
well studied in translational invariant systems, but less considered in a
nano-conductor. We make connections between electron/phonon drag and other
related effects, e.g., the current-induced nonconservative, effective
magnetic force\cite{DuMcTo.2009,JMP.2010,BoKuEgVo.2011}, and adiabatic pumping in a coherent conductor\cite{Buttiker94,Brouwer98}. These effects can only emerge
in a nanoconductor with at least two phonon degrees of freedom. This makes our study different and complementary 
to most of other works\cite{entin-wohlman_three-terminal_2010,leijnse_nonlinear_2010,SanBu11,Ren2012,Zhou2015}.
Furthermore, we extend the analysis to the nonlinear
regime, look at the drag effect on energy transfer between electrons and
phonons, and discuss the possibility of driving heat flow using an electric
device, or charge transport using
heat\cite{entin-wohlman_three-terminal_2010,SanBu11}.

The paper is organized as follows. In Sec.~\ref{sec:loe}, we introduce our
model setup, and present analytical results for the charge and heat currents
focusing on the electron/phonon drag effect. In Sec.~\ref{sec:num} we analyze a
simple one-dimensional (1D) model system to illustrate that the drag effect
shares the same origin as that in a translational invariant lattice, and can be
understood as a result of the momentum transfer between electrons and phonons.
We also provide numerical results for the model system.  Section~\ref{sec:conc}
gives concluding remarks. Finally, the details of the derivation
are given in Appendices~\ref{sec:ephcof}-\ref{sec:eew}.

%%%%%%%%%%%%%%%%%%%%%%%%%%%%%%%%%%%%%%%%%%%%%%%%%%%%%%%%%%%%%%%%%%%%%%%%%%%%%%%%%%%%%%%
%\section{Model and methods}
%\label{sec:negf}
%%%%%%%%%%%%%%%%%%%%%%%%%%%%%%%%%%%%%%%%%%%%%%%%%%%%%%%%%%%%%%%%%%%%%%%%%%%%%%%%%%%%%%%
\section{Thermoelectric transport}
\label{sec:loe}
\subsection{System setup and Hamiltonian}
\begin{center}
\begin{figure}[h]
\includegraphics[scale=0.5]{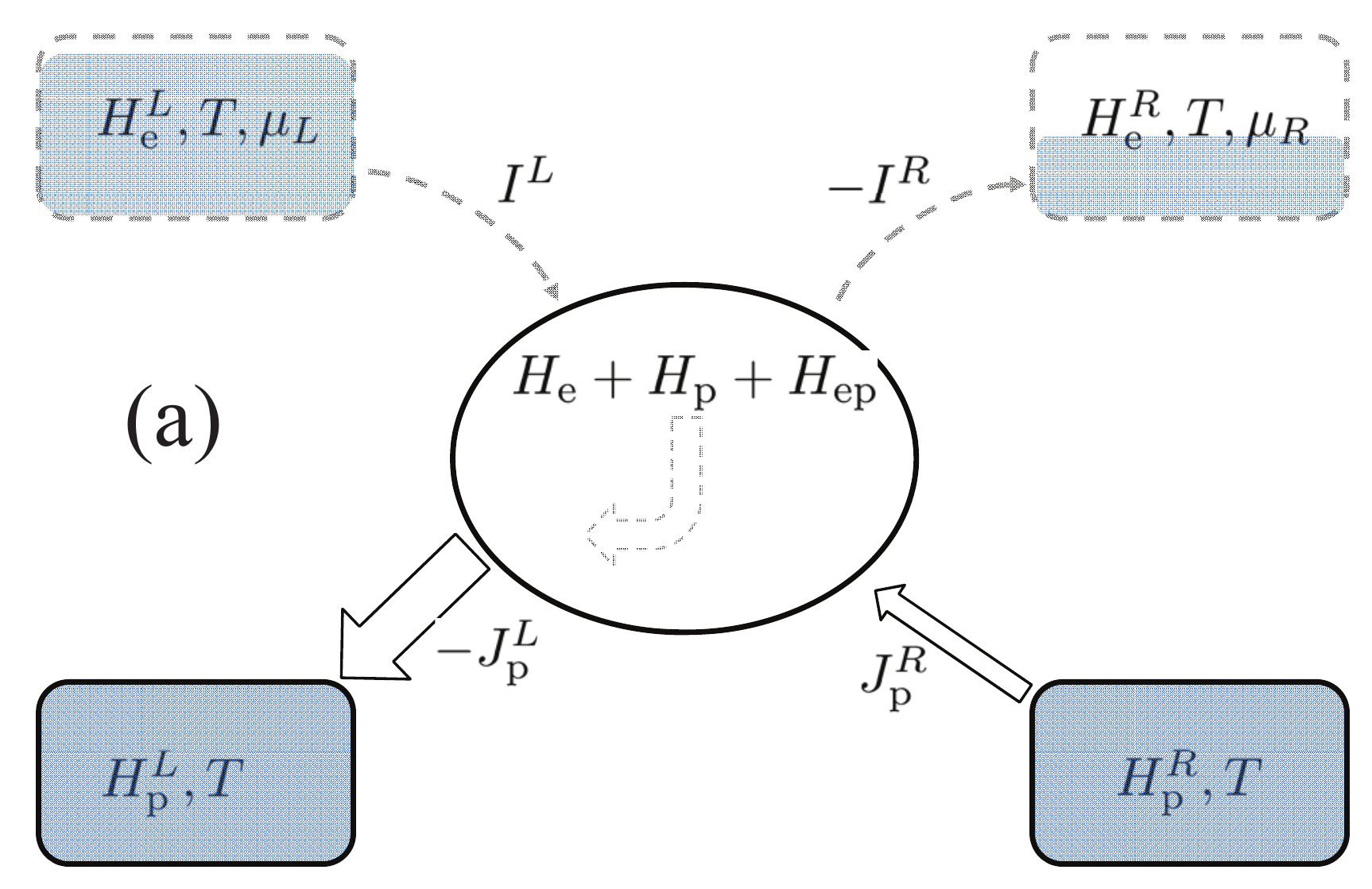}\\
\includegraphics[scale=0.5]{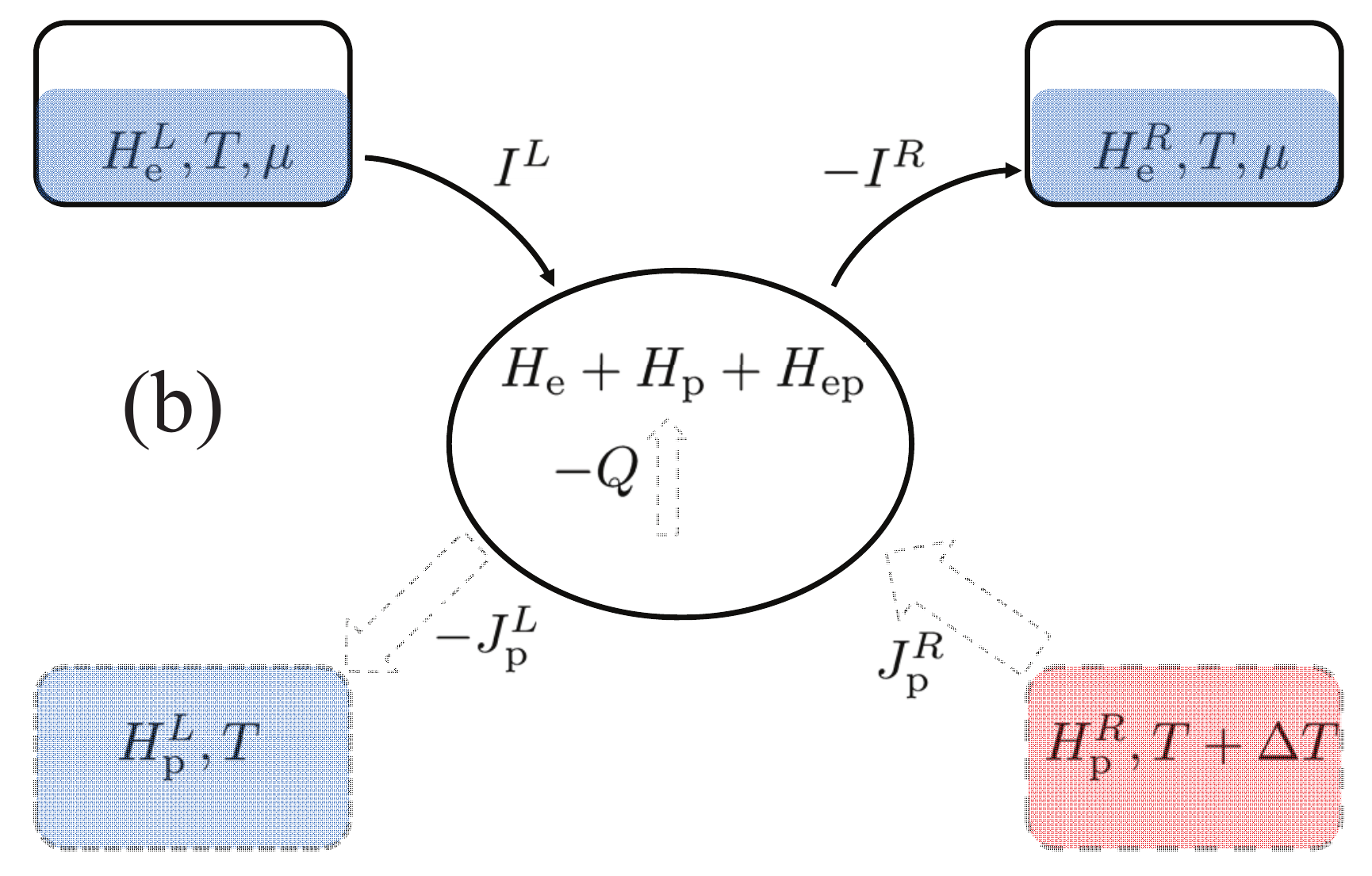}
\caption{Schematic representation of the model system considered in this paper, and its different possible situations:
(a) By applying a voltage bias, heat can be extracted from one of the
phonon baths, although they remain at the same temperature. 
%The electrical and
%heat current are defined to be positive in the direction from the bath to the
%device. The power is positive in the direction from electrons to phonons.
%(b) At zero voltage bias, the two electron baths are kept at the same
%temperature, but different from that of phonon baths. Besides heat transport between
%electrons and phonons, there could also be an electronic current between the
%electron baths.
(b) By applying a temperature difference between the two phonon baths, an electrical current can be generated between the two electron baths. This contributes with a phonon-drag part to the thermopower. 
} 
\label{fig:sys}
\end{figure}
\end{center}

We consider a model device containing an electronic ($H_{\rm e}$) and a
phononic (vibrational) part ($H_{\rm p}$), with interactions between them
$H_{\rm ep}$. The electronic part is linearly coupled to two separate electron
baths ($L, R$), so does the vibrational part (Fig.~\ref{fig:sys}). The coupling matrix
is denoted by $V_\sigma^\alpha$. The Hamiltonian of the entire system is written as
\begin{equation}
	H = \sum_{\sigma={\rm e,p}}H_\sigma + H_{\rm ep} + \sum_{\alpha=L,R;\sigma={\rm e,p}}(H_\sigma^{\alpha}+V_\sigma^{\alpha}).
	\label{eq:h}
\end{equation}
The electron and phonon subsystem (device plus left and right baths) are
noninteracting. For example, the phonons are described within the harmonic
approximation, and the electrons within a single particle picture. We assume no
direct coupling between these baths.  The only many-body interaction is 
$H_{\rm ep}$ within the device. In a
tight-binding description of the electronic Hamiltonian, it can be written as
\begin{equation}
	H_{\rm ep} = \sum_{i,j,k}M_{ij}^k (c_i^\dagger c_j + {\rm h.c.}){u}_k,
	\label{eq:eph}
\end{equation}
where $c_i^\dagger$($c_j$) is the electron creation (annihilation) operator for
the $i$-($j$-)th electronic site, and ${u}_k$ is the mass-normalized
displacement away from the equilibrium position of the $k$-th degrees of
freedom, i.e., $u_k = \sqrt{m_k} r_k$, with $m_k$ the mass of the $k$-th degree
of freedom, and $r_k$ its displacement away from equilibrium position.
$M_{ij}^k$ is the e-ph interaction matrix element.

%The separation of the electron and phonon baths makes the theoretical
%development easier. In reality, they could either be physically separated, or
%built into one.  For example, one electrode could serve both as an
%electron and a phonon bath.  But we assume that we have independent control
%over their temperatures $T^\alpha_{\rm e}$ and $T^\alpha_{\rm p}$, $\alpha=L,R$.
%
To calculate the electrical and heat current, we assume the
e-ph interaction is weak, and keep only the lowest order
self-energies\cite{Lu2015Rev}. We perform an expansion of the Green's functions and current up to the second order in $M$, following the idea of
Ref.~\onlinecite{LOE1}.  For example, the lesser Green's function is expanded as
\begin{eqnarray}
	G^< 	\approx G^<_0+G^r_0\Sigma^<_{\rm ep}G^a_0
	\!\!&+&\!\!G^r_0\Sigma_{\rm ep}^rG^r_0\Sigma^<_{}G^a_0\nonumber\\
	\!\!&+&\!\!G^r_0\Sigma^<_{}G^a_0\Sigma^a_{\rm ep}G_0^a,
	\label{eq:loe1}
\end{eqnarray}
with $\Sigma_{}=\Sigma_L+\Sigma_R$ the self-energy due to coupling to
electrodes, $\Sigma_{\rm ep}$ self-energy due to e-ph interaction, and $G_0$
the noninteracting electron Green's function. Similar expression holds for
$G^>$ and $D^{>,<}$. The details of the method can be found in
Refs.~\onlinecite{LOE1,Lu07,Lu2015Rev}.
%We have listed the expressions for the current in Appendix~\ref{sec:negf},
%for details see Refs.~\onlinecite{LOE1,Lu07,Lu2015Rev}.

%-----------------------------------------------------
\subsection{Linear transport coefficients}
\label{subsec:linear}
In the linear response regime, we introduce an infinitesimal change of the chemical
potential or temperature at one of the baths, $\alpha$, e.g., $\mu^\alpha=\mu+\delta \mu$,
$T_\sigma^\alpha=T+\delta T_\sigma$, with $\mu$ and $T$ the corresponding
equilibrium values.  We look at the response of the charge and heat current due
to this small perturbation. Up to the 2nd order in $M$, the result is summarized as follows
\begin{equation}
	\left(
	\begin{array}{ccc}
		\frac{I^\alpha}{e}\\
		J^\alpha_{\rm e}\\
		J^\alpha_{\rm p}
	\end{array}
	\right)
	=
	\left(
	\begin{array}{ccc}
		\mathcal{L}_0 & \mathcal{L}_1 & \tilde{\mathcal{Q}}_0\\
		\mathcal{L}_1 & \mathcal{L}_2 & \tilde{\mathcal{Q}}_1\\
		{\mathcal{Q}}_0&{\mathcal{Q}}_1&\mathcal{K}_{\rm p}\\
	\end{array}
	\right)
	\left(
	\begin{array}{ccc}
		\delta \mu\\
		\frac{\delta T_{\rm e}}{T}\\
		\frac{\delta T_{\rm p}}{T}
	\end{array}
	\right).
	\label{eq:lang2}
\end{equation}
We define the positive current direction as that electrons/phonons go from the
bath to the device, and $I^\alpha$, $J^\alpha_{\rm e}$, $J^\alpha_{\rm p}$ are
the electrical current, heat current carried by electrons and phonons,
respectively. The expressions for the coefficients $\mathcal{L}$ and
$\mathcal{K}_{\rm p}$ are given in Appendix \ref{sec:ephcof}. Both include three
contributions. The first term is the elastic Landauer result. The second term
is the (quasi-)elastic correction due to change of the electron spectral
function.  The last one is the inelastic term. The effect of each part in
$\mathcal{L}$ on the conductance and Seebeck coefficient have been analyzed in
Ref.~\onlinecite{Lu2015Rev} for a single level model. $\mathcal{Q}_n$ and
$\tilde{\mathcal{Q}}_n$ are the drag coefficients. We use the following convention
for the drag effect: The electron drag effect
corresponds to generating phonon flow due to electron flow, while the phonon drag corresponds
to the opposite process. We can write $\mathcal{Q}_n$ and $\tilde{\mathcal{Q}}_n$ as
\begin{eqnarray}
	\tilde{\mathcal{Q}}_n\!\! &=&\!\! -\!\!\sum_{\beta}\!\! \int\!\! \frac{d\omega}{2\pi}\hbar\omega\textrm{Tr}[\Lambda^{(n)}_{\tilde{\alpha}\beta}(\omega)\mathcal{A}_\alpha(\omega)]\partial_{\hbar\omega} n_B(\hbar\omega,T),\\
	\label{eq:corQ}
	{\mathcal{Q}}_n\!\! &=& \!\!-\!\!\sum_{\beta}\!\! \int\!\! \frac{d\omega}{2\pi}\hbar\omega\textrm{Tr}[\Lambda^{(n)}_{\alpha\beta}(\omega)\tilde{\mathcal{A}}_{\alpha}(\omega)]\partial_{\hbar\omega} n_B(\hbar\omega,T),
	\label{eq:cortQ}
\end{eqnarray}
where $n_B(\hbar\omega,T)=\left[{\rm exp}\left(\frac{\hbar\omega}{k_BT}\right)-1\right]^{-1}$ is the Bose-Einstein distribution function.  Throughout the paper, we use ${\rm Tr}[\cdot]$ for
trace over phonon indices, ${\rm tr}[\cdot]$ for trace over electronic
degrees of freedom, $\tilde{\mathcal{A}}_\alpha/\mathcal{A}_\alpha$ is the
(time-reversed) phonon spectral function [Eq.~(\ref{eq:pha})], and
$\Lambda^{(n)}_{\tilde{\alpha}\beta}(\omega)$ is defined as
\begin{eqnarray}
	\label{eq:lamab}\Lambda^{(n)}_{\tilde{\alpha}\beta}(\omega) &=& \int \frac{d\varepsilon}{2\pi}(\varepsilon-\mu_0)^nX_{\tilde{\alpha}\beta}(\varepsilon,\varepsilon_-)\\
	&\times&\bigl[f(\varepsilon,\mu_\alpha,T_\alpha^e)-f(\varepsilon_-,\mu_\beta,T_\beta^e)\bigr],\nonumber\\
	X_{\tilde{\alpha}\beta}(\varepsilon,\varepsilon_-) &=& \textrm{tr}\bigl[M\tilde{A}_\alpha(\varepsilon) MA_{\beta}(\varepsilon_-)\bigr], \label{eq:X}
\end{eqnarray}
with $\varepsilon_-=\varepsilon-\hbar\omega$, and $\mu_0$ the equilibrium chemical potential. In the definition of $X_{\tilde{\alpha}\beta}$ and $\Lambda^{(n)}_{\tilde{\alpha}\beta}$, $\tilde{\alpha}$ means we need
to use the time-reversed electron spectral function $\tilde{A}_\alpha$ [Eq.~(\ref{eq:ea})].
$\Lambda^{(0)}(\omega)$ is the coupling-weighted electron-hole pair density of states (DOS), 
introduced in our previous work\cite{JMP.2010,luprb12,Lu15}.
In the linear regime, the Fermi distribution
$f(\varepsilon,\mu_\alpha,T^\alpha)=f(\varepsilon,\mu,T)$ is the same for both
electrodes, with $f(\varepsilon,\mu,T)=\left[{\rm exp}\left({\frac{\varepsilon-\mu}{k_BT}}\right)+1\right]^{-1}$.
Hereafter, the summation of $\beta$ is over $L$ and $R$, 
and the integration is from $-\infty$ to $+\infty$ if not specified explicitly.
%We have defined \revision{define nonequilibrium $\Lambda$ here??}
%\begin{eqnarray}
%	%\label{eq:lamdaa}\Lambda^{(n)}(\omega)&=&\sum_{\alpha,\beta}\Lambda^{(n)}_{\tilde{\alpha}\beta}(\omega),\\ 
%	\label{eq:lamab}\Lambda^{(n)}_{\tilde{\alpha}\beta}(\omega) &=& \int \frac{d\varepsilon}{2\pi}(\varepsilon-\mu)^nX_{\tilde{\alpha}\beta}(\varepsilon,\varepsilon_-)\\
%	&\times&\bigl[f(\varepsilon,\mu_\alpha,T_\alpha^e)-f(\varepsilon_-,\mu_\beta,T_\beta^e)\bigr],\nonumber\\
%	X_{\tilde{\alpha}\beta}(\varepsilon,\varepsilon_-) &=& \textrm{tr}\bigl[M\tilde{A}_\alpha(\varepsilon) MA_{\beta}(\varepsilon_-)\bigr], \label{eq:X}
%\end{eqnarray}
%with $\varepsilon_-=\varepsilon-\hbar\omega$, $f(\varepsilon,\mu,T)=\left[{\rm exp}\left({\frac{\varepsilon-\mu}{k_BT}}\right)+1\right]^{-1}$ the Fermi-Dirac distribution, and  $n(\hbar\omega,T)=\left[{\rm exp}\left(\frac{\hbar\omega}{k_BT}\right)-1\right]^{-1}$ the Bose-Einstein distribution. Here, ${\rm Tr}\{\cdot\}$ is over the electronic degrees of freedom. In the definition of $X_{\tilde{\alpha}\beta}$ and $\Lambda^{(n)}_{\tilde{\alpha}\beta}$, $\tilde{\alpha}$ means we need
%to use the time-reversed spectral function $\tilde{A}_\alpha$ [Eq.~(\ref{eq:ea})].
%$\Lambda^{(0)}(\omega)$ is the coupling-weighted electron-hole pair density of states (DOS), 
%introduced in our previous work\cite{JMP.2010,luprb12,Lu15}.
In the linear regime, without magnetic field, we have $(D^r)^T=D^r$ and
$(G^r)^T=G^r$. This leads to $\mathcal{Q}_n=\tilde{\mathcal{Q}}_n$, which
ensures the Onsager symmetry (Appendix \ref{sec:ons}).

For one electronic level coupled to one phonon mode, we can check that our
result for $\mathcal{Q}_0/{\mathcal{Q}}_1$ is equivalent to that of
Ref.~\onlinecite{entin-wohlman_three-terminal_2010}(Appendix \ref{sec:eew}).
In order for $\mathcal{Q}_0/{\mathcal{Q}}_1$ to be non-zero, we need some
special design of the system, e.g., asymmetric coupling to the left and right
electron bath\cite{entin-wohlman_three-terminal_2010}.

Here, we focus on the case where there are two or more phonon modes.  For
${\tilde{\mathcal{Q}}}_0$, we can do an expansion over the energy dependence of
the electron spectral function. The zeroth order contribution is
\begin{eqnarray}
	\label{eq:corQwb}
	\tilde{\mathcal{Q}}_0^{(0)} &=& \sum_\beta \int \frac{d\omega}{4\pi^2} (\hbar\omega)^2 \partial_{\hbar\omega} n_B(\hbar\omega,T)\\
	&\times&\textrm{Tr}[{\rm Im} X_{\tilde{\alpha}\beta}(\mu,\mu){\rm Im}{\mathcal{A}}_\alpha(\omega)]\nonumber,
\end{eqnarray}
with Re and Im meaning real and imaginary part, respectively. 
%The phonon spectral function $\mathcal{A}_\alpha = D_0^r\Pi_\alpha D_0^a$, and 
%\begin{equation}
%	\label{eq:x2}X_{\tilde{\alpha}\beta}(\varepsilon,\varepsilon-\hbar\omega) = \textrm{Tr}\left[M\tilde{A}_\alpha(\varepsilon) MA_{\beta}(\varepsilon-\hbar\omega)\right].
%\end{equation}
%$n(\hbar\omega,T)$ is the Bose distribution function (Eq.~\ref{eq:be}). 
In order for $\tilde{\mathcal{Q}}_0^{(0)}$ to be non-zero, the device needs to have at least
two vibrational modes. This follows from the fact that $\mathcal{A}_\alpha  $
is Hermitian.  

%The first order contribution to $\tilde{\mathcal{Q}}_0$ is 
%\begin{eqnarray}
%	\tilde{\mathcal{Q}}^{(1)}_0 &=& \sum_\beta \int \frac{d\omega}{4\pi^2} \frac{(\hbar\omega)^3}{2} \partial_{\hbar\omega} n(\hbar\omega,T)\\
%	&\times&\textrm{Tr}\left[{\rm Re}\mathcal{A}_\alpha(\omega){\rm Re}\bigl(X_{\tilde{\alpha}'{\beta}}(\mu,\mu)- X_{\tilde{\alpha}{\beta'}}(\mu,\mu)\bigr)\right].\nonumber
%	\label{eq:corQwb1}
%\end{eqnarray}
%with
%\begin{equation}
%	X_{\tilde{\alpha}'{\beta}}(\mu,\mu) = {\rm tr}\bigl[ M \partial_\varepsilon \tilde{A}_{\alpha}(\mu)M{A}_{\beta}(\mu) \bigr].
%	\label{eq:xx}
%\end{equation}
%

\subsubsection{Relation with adiabatic pumping and current-induced forces}
Now we write $\tilde{\mathcal{Q}}_0^{(0)}$ in terms of the unperturbed retarded (advanced) electron scattering states, coming from (leaving to) the
left $|\psi_L\rangle$ ($|\tilde{\psi}_L\rangle$)  or right $|\psi_R\rangle$  ($|\tilde{\psi}_R\rangle$) electrode
\begin{eqnarray}
	\label{eq:xkl}
	X^{kl}_{\tilde{\alpha}\beta}(\varepsilon,\varepsilon_-)
	&=& \sum_{m,n} \langle \psi_{\beta}^n(\varepsilon_-)|M^k|\tilde{\psi}_{\alpha}^m(\varepsilon)\rangle 
	\langle \tilde{\psi}^m_{\alpha}(\varepsilon)|M^l|\psi_{\beta}^n(\varepsilon_-)\rangle.\nonumber\\
\end{eqnarray}
%\revision{According to the Lippmann-Schwinger equation}, 
Here, $m$ and $n$ are channel indices.
The retarded and advanced scattering states including e-ph interaction are generated from 
\begin{eqnarray}
	\label{eq:lip}
	|\Psi_\alpha(\varepsilon)\rangle &=& |\psi_\alpha(\varepsilon)\rangle + G^{r} H_{\rm ep} |\psi_\alpha(\varepsilon)\rangle,\\
	|\tilde{\Psi}_\alpha(\varepsilon)\rangle &=& |\psi_\alpha(\varepsilon)\rangle + G^{a} H_{\rm ep} |\psi_\alpha(\varepsilon)\rangle,
	\label{eq:sch}
\end{eqnarray}
They are normalized as
\begin{equation}
	\langle \Psi_\alpha(\varepsilon)|\Psi_\beta(\varepsilon')\rangle = 2\pi \delta(\alpha,\beta)\delta(\varepsilon-\varepsilon'),
	\label{}
\end{equation}
and so is $|\psi_\alpha\rangle$.
From the definition of the scattering matrix
\begin{equation}
2\pi\delta(\varepsilon-\varepsilon')S_{\alpha\beta}^{mn}=\langle \tilde{\Psi}_\alpha^m(\varepsilon')|\Psi_\beta^n(\varepsilon)\rangle,
	\label{}
\end{equation}
and Eqs.~(\ref{eq:lip}-\ref{eq:sch}),
we get
\begin{equation}
	S_{\alpha\beta}^{mn} = \delta_{\alpha,\beta}\delta_{m,n} - i \langle {\psi}_\alpha^m|H_{\rm ep}|\Psi_\beta^n\rangle.
	\label{}
\end{equation}
Here, $S_{\alpha\beta}^{mn}$ is the matrix element connecting the incoming wave
from the $n$-th channel in electrode $\beta$ to the $m$-th outgoing channel in electrode $\alpha$. Taking the derivative over the phonon displacement yields 
\begin{equation}
	\partial_k S_{\alpha\beta}^{mn} = - i \langle \tilde{\Psi}_\alpha^m|M^k|\Psi_\beta^n\rangle.
	\label{eq:sm}
\end{equation}
Substituting Eq.~(\ref{eq:sm}) into Eq.~(\ref{eq:xkl}), taking the $\omega\to 0$ limit, we obtain
\begin{equation}
	{\rm Im} \sum_\beta {\rm tr}\bigl[\partial_l S_{\alpha\beta} \partial_k S_{\beta\alpha}^\dagger\bigr]={\rm Im} \sum_\beta X_{\tilde{\alpha}\beta}^{kl}+\cdots.
	\label{eq:brouwer}
\end{equation}
The trace is over the channel indices. We have kept only the second order terms. 

Equation~(\ref{eq:brouwer}) makes
connection with the Brouwer formula for adiabatic pumping\cite{Brouwer98,Thomas2012,Bustos13}.
Here, a temperature difference between the left and right electrode breaks
the population balance between the phonon scattering states from these two
baths, e.g., there are more phonon waves travelling in one direction, determined by
the temperature bias. When the phonon wave goes through the device, it produces
phase-shifted oscillating potential felt by the electrons. In the space of the atomic
coordinates, the trajectory may form a closed loop, generating pumped
electrical current.

The opposite of this effect is that an electrical current generates a directed
phonon heat current. The term governing this effect is $X_{LR}$. The same term appears
in the expressions for the current-induced nonconservative and effective
magnetic forces\cite{DuMcTo.2009,JMP.2010,BoKuEgVo.2011,tnt-belstein,luprb12},
e.g., Eqs.~(56-61) in Ref.~\onlinecite{luprb12}. This shows that the electron
drag effect is closely related to these novel current-induced forces.

\subsection{Nonlinear regime}
When the applied temperature or voltage bias is
large, additional energy transfer between the electron and phonon subsystem takes place.
We consider two situations: Electrical-current-driven heat flow in the isothermal case, and temperature-driven electrical current at zero voltage bias.

\subsubsection{Electrical current induced heat flow\\ ($T^{e}=T^{\rm p}=T$, $eV \neq 0$)}
\begin{figure}[htpb]
	\begin{center}
		\includegraphics[scale=0.4]{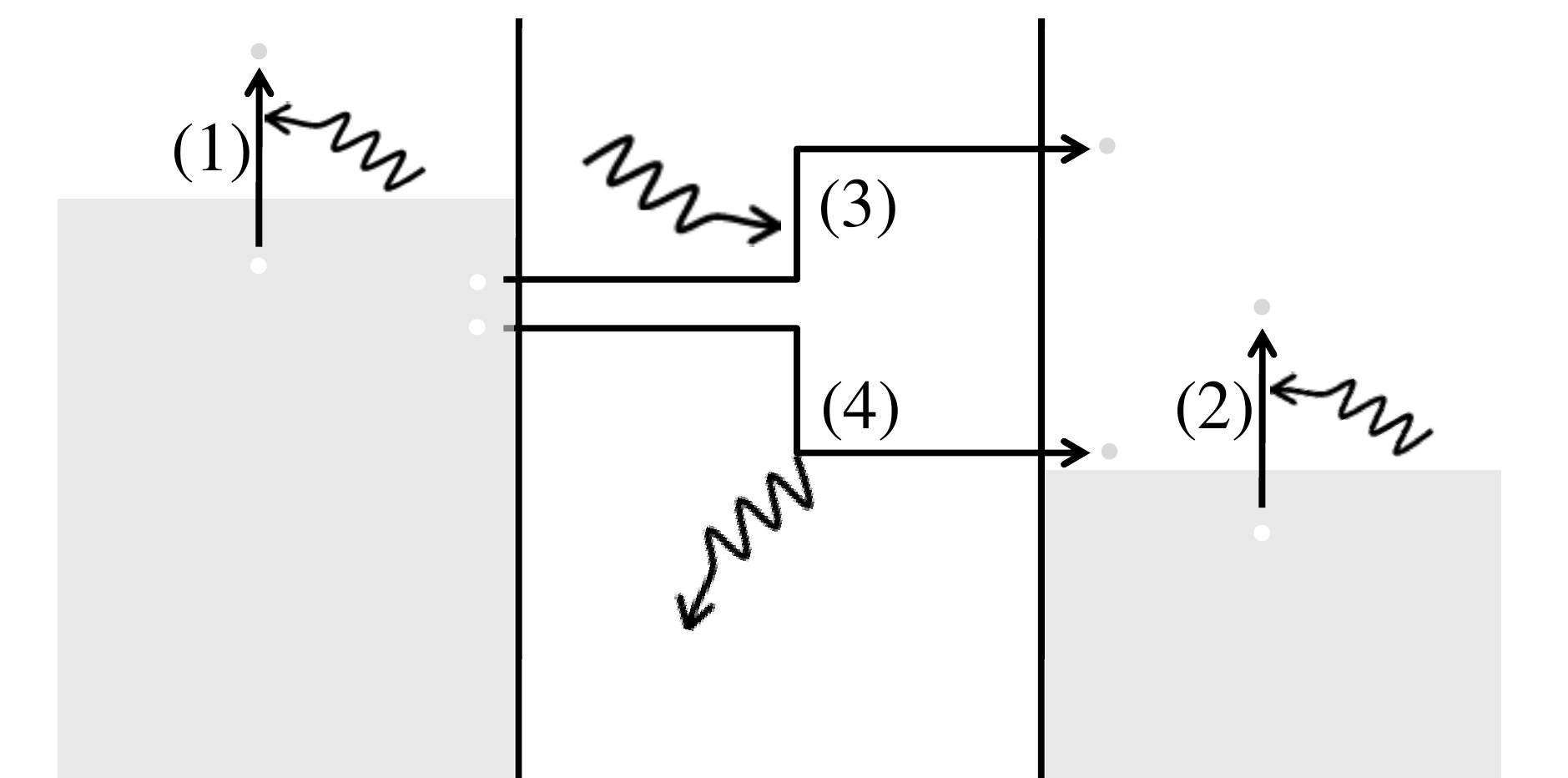}
	\end{center}
	\caption{Electron-hole pair excitation and recombination processes. (1) and (2) are intra-electrode processes. (3) and (4) are inter-electrode ones. At $T=0$, (1)-(3) are not possible, and (4) is possible only when the applied bias is larger than the phonon energy.}
	\label{fig:ehpair}
\end{figure}

In the first setup, all the baths are at the same temperature ($T$), but the electron baths are subject to a nonzero voltage bias ($eV=\mu_L-\mu_R$). This is the
most common situation in a working electronic device [Fig.~\ref{fig:sys} (a)]. For large bias, there will be energy transfer from the electron to the phonon
subsystem
\begin{eqnarray}
	Q\!&=&\!\! 2 \sum_{\alpha,\beta} \int\frac{d\omega}{2\pi} \int \frac{d\varepsilon}{2\pi} \hbar\omega {\rm Tr}\left[{\rm tr}\left[ MA_\alpha(\varepsilon)MA_\beta(\varepsilon+\hbar\omega) \right]\mathcal{A}(\omega)\right]\nonumber\\
	&&\times f_\beta(\varepsilon+\hbar\omega)(1-f_\alpha(\varepsilon))(n_B(\hbar\omega,T)+1).
	\label{eq:power0}
\end{eqnarray}
Since all the baths are at the same temperature, we have omitted it in this
subsection. Equation~(\ref{eq:power0}) is a result of
balance between phonon emission and adsorption processes [Fig.~\ref{fig:ehpair}]. 
For $\omega >0$, it
represents process where an electron  in electrode $\beta$ combines with a hole
at lower energy in $\alpha$, accompanied by a phonon emission process. For
$\omega<0$, it represents the opposite process, where an electron-hole pair is
created between $\alpha$ and $\beta$ by adsorbing one phonon. While the Fermi distributions
$f_\beta(1-f_\alpha)$ ensures that the phonon emission process happens only
when the applied bias $eV$ is larger than the phonon energy $\hbar\omega$, the
Bose function $n_B+1$ prohibits phonon adsorption process at $T=0$.
This equation can also be written in a compact form as
\begin{eqnarray}
	Q\!&=&\!\!\int \frac{d\omega}{2\pi} \hbar\omega {\rm Tr}\bigl[\Lambda^{(0)}_{LR}(\omega){\mathcal{A}}(\omega)\bigr] \Delta n_B(\hbar\omega,T;\hbar\omega-eV,T),\nonumber\\
	\label{eq:power}
\end{eqnarray}
with
\begin{equation}
	\Delta n_B(\hbar\omega_1,T_1;\hbar\omega_2,T_2)=n_B(\hbar\omega_1,T_1)-n_B(\hbar\omega_2,T_2).	
	\label{eq:dn}
\end{equation}
Energy transfer within the device breaks the balance between the device and bath phonons. As a result, the extra energy is further transferred to the two
phonon baths. The heat current flowing out of the phonon bath $\alpha$ is
given by the minus of Eq.~(\ref{eq:power}) with $\tilde{\mathcal{A}}$ replaced
by $\tilde{\mathcal{A}}_\alpha$, such that
%\begin{eqnarray}
%	J_{\rm p}^{\alpha}\!&=&\!-\!\int \frac{d\omega}{2\pi} \hbar\omega {\rm Tr}\left[\Lambda^{(0)}_{LR}(\omega)\tilde{\mathcal{A}}_{\alpha}(\omega)\right]\Delta n(\hbar\omega;\hbar\omega-eV).\nonumber\\ \label{eq:pphL2}
%\end{eqnarray}
%Comparing with Eq.~(\ref{eq:power}), we see
\begin{equation}
	J_{\rm p}^{L}+J_{\rm p}^{R}+Q=0,
	\label{eq:econs}
\end{equation}
as required by the energy conservation.

%Now we provide one intuitive way to understand these results. Comparing
%Eqs.~(\ref{eq:power}-\ref{eq:pphL2}) with the Landauer formula for phonon heat
%current Eq.~(\ref{eq:landauer3}), we can see that $\Lambda^0(\omega)$ plays the role of $\Gamma^{\rm p}_\alpha$ for an equilibrium phonon bath.
%So we can define an \emph{current-induced} effective transmission from the electron to one of the phonon bath $\alpha$ as
%\begin{equation}
%  \mathcal{T}_{n,\alpha}(\omega) = {\rm Tr}\left[ \Lambda^0_{LR}(\omega) \tilde{\mathcal{A}}_{\alpha}(\omega) \right].
%	\label{eq:etel}
%\end{equation}
%Since the electron subsystem breaks the time-reversal symmetry (by the bias
%voltage), the effective coupling $\Lambda^0_{LR}(\omega)$ has an imaginary
%part. This is the key difference between an equilibrium bath and a
%nonequilibrium bath, and is responsible for the mutual drag between electrons and
%phonons.  The effective transmission due to the imaginary part of
%$\Lambda^0_{LR}$ is positive for one phonon bath, and negative for the other.
%This follows from $\Im \tilde{\mathcal{A}}=\Im \tilde{\mathcal{A}}_L +\Im \tilde{\mathcal{A}}_R=0$.
%

In fact, we can split $J_{\rm p}^\alpha$ into two parts according to their
symmetry upon bias reversal
$J_{\rm p}^\alpha=J_{\rm p}^{\alpha,h}+J_{\rm p}^{\alpha,p}$, where  $J_{\rm p}^{\alpha,h}$ and
$J_{\rm p}^{\alpha,p}$ are even and odd function of $eV$.  We call them the Joule heating
and Peltier drag current, respectively.  Assuming constant electron DOS, we get
\begin{eqnarray}
	J_{\rm p}^{\alpha,h}\!\!&\approx&\!\!-\!\!\int_{0}^{+\infty} \frac{d\omega}{4\pi^2}\; h(\hbar\omega){\rm{Tr}}[{\rm Re}X_{LR}(\mu,\mu) {\rm Re}\tilde{\mathcal{A}}_\alpha(\omega)],\nonumber\\
	\label{eq:wbppL1}\\
	J_{\rm p}^{\alpha,p}\!\!&\approx&\!\!-\!\!\int_{0}^{+\infty} \frac{d\omega}{4\pi^2}\; p(\hbar\omega){\rm Tr}[{\rm Im}X_{LR}(\mu,\mu){\rm Im}\tilde{\mathcal{A}}_\alpha(\omega)].\nonumber\\
	\label{eq:wbppL2}
\end{eqnarray}
The two coefficients are
\begin{eqnarray}
	h(\hbar\omega)\!\! &\equiv&\!\! \sum_{s=\pm 1}\!\!\hbar\omega(\hbar\omega+s\,eV)\Delta n_B(\hbar\omega+s eV;\hbar\omega),\\
	p(\hbar\omega)\!\! &\equiv&\!\! \sum_{s=\pm 1}\!\!s\,\hbar\omega(\hbar\omega+s\,eV)\Delta n_B(\hbar\omega+s eV;\hbar\omega).
	\label{eq:hp}
\end{eqnarray}
The Joule current corresponds to the 
energy transfer from the electrons to the phonons in Eq.~(\ref{eq:power}), i.e., $J_{\rm p}^{L,h}+J_{\rm p}^{R,h}+Q=0$.
But the drag current is related to the $\mathcal{Q}_0$ coefficient in Subsec.~\ref{subsec:linear}, and depends on the direction of current flow,
i.e., $J_{\rm p}^{L,p}+J_{\rm p}^{R,p}=0$. This relation follows from the fact that ${\rm Im} \mathcal{A} ={\rm Im} \tilde{\mathcal{A}} ={\rm Im} \tilde{\mathcal{A}}_L + {\rm Im} \tilde{\mathcal{A}}_R = 0$.  We will see later that it is due to momentum transfer between electrons and phonons.
In the limit of high temperature ($k_BT\gg eV \pm \hbar\omega$), we have
$h(\hbar\omega)\to 0$, and $p(\hbar\omega)\to 2eV k_BT$. The drag part will
dominate over the Joule heating part. In this case, it is possible to extract
heat from one of the phonon baths by applying a voltage bias, similar to a
refrigerator, as shown in Fig.~\ref{fig:sys} (b).

We note that in Ref.~\onlinecite{Lu15}, we have studied
the same problem using the semi-classical generalized Langevin equation approach. 
Similar equations were derived there, and the asymmetric heat flow was attributed
to the asymmetric current-induced forces. These two complementary analysis shows
that the two effects are closely related.

\subsubsection{Temperature-driven electric current\\ ($\mu_L=\mu_R$,$T^e\neq T^{\rm p}$)} 
In the second setup, we apply a temperature difference
between the electron and phonon subsystem at zero voltage bias.  This drives an electrical current
within the device
\begin{eqnarray}
	{I}^\alpha=e\sum_{\beta}\int \frac{d\omega}{2\pi} {\textrm{Tr}}\bigl[\Lambda^{(0)}_{\tilde{\alpha}\tilde{\bar{\alpha}}}(\omega)\mathcal{A}_\beta(\omega)\bigr]
	\Delta n_B(\hbar\omega,T_{\rm p}^\beta;\hbar\omega,T_{\rm e}).\nonumber\\
	\label{eq:i4r}
\end{eqnarray}
Here, $\bar{\alpha}$ means the lead different from $\alpha$.
There are two possible situations here. The first one is that the phonon baths are at the same temperature ($T_{\rm p}$), but different from that of electron baths ($T_{\rm e}$). We can consider the two phonon baths as an effective single bath. The four-terminal setup reduces to a three-terminal one, and equation~(\ref{eq:i4r}) simplifies to
\begin{eqnarray}
	{I}^\alpha=e\int \frac{d\omega}{2\pi} {\textrm{Tr}}\left[\Lambda^{(0)}_{\tilde{\alpha}\tilde{\bar{\alpha}}}(\omega)\mathcal{A}(\omega)\right]
	\Delta n_B(\hbar\omega,T_{\rm p};\hbar\omega,T_{\rm e}).
	\label{eq:i4r1}
\end{eqnarray}
The three-terminal setup has been considered in
Ref.~\onlinecite{entin-wohlman_three-terminal_2010}. For a single electronic level coupling to one phonon mode, equation~(\ref{eq:i4r1}) agrees with result therein.  Due to the temperature difference between the electron and phonon systems, there will be energy flow between them. It has been analyzed in Sec. III C of
Ref.~\onlinecite{Lu2015Rev}. A similar problem has been considered in Refs.~\onlinecite{segal_single_2008,Lifa13,renjie13,Vinkler2014}. Here we focus on the other situation, where we
apply a temperature difference between the two phonon baths [Fig.~\ref{fig:sys}
(b)]. This generates a phonon-drag electrical current. For constant electronic
DOS, we get
\begin{eqnarray}
	{I}^\alpha\approx e\int \frac{d\omega}{4\pi^2} &\hbar\omega &{\textrm{Tr}}\left[{\rm Im} X_{\tilde{\alpha}\tilde{\bar{\alpha}}}(\mu,\mu){\rm Im}\mathcal{A}_{\bar{\alpha}}(\omega)\right]\nonumber\\
	&\times&\Delta n_B(\hbar\omega,T_{\rm p}^\alpha;\hbar\omega,T_{\rm p}^{\bar{\alpha}}),
	\label{eq:i4rwb}
\end{eqnarray}
extending the result in Subsec.~\ref{subsec:linear} to the nonlinear regime.

%\begin{figure}[!htbp]
%\includegraphics[scale=0.4]{system5.pdf}
%\caption{At zero voltage bias, the two electron baths are kept at the same
%temperature, but different from that of phonon baths. Besides heat transport between
%electrons and phonons, there could also be an electronic current between the
%electron baths.} 
%\label{fig:sys2}
%\end{figure}

%\begin{figure}[!htbp]
%\includegraphics[scale=0.4]{system4.pdf}
%\caption{By applying a temperature difference between the two phonon baths, an electrical current can be generated between the two electron baths. This contributes a phonon-drag part to the thermopower.} 
%\label{fig:sys4}
%\end{figure}

%-----------------------------------------------------
%\section{Numerical results}
%-----------------------------------------------------

\section{Model Calculation}
\label{sec:num}
\subsection{1D model and qualitative analysis}
\begin{figure}[h]
\includegraphics[scale=0.35]{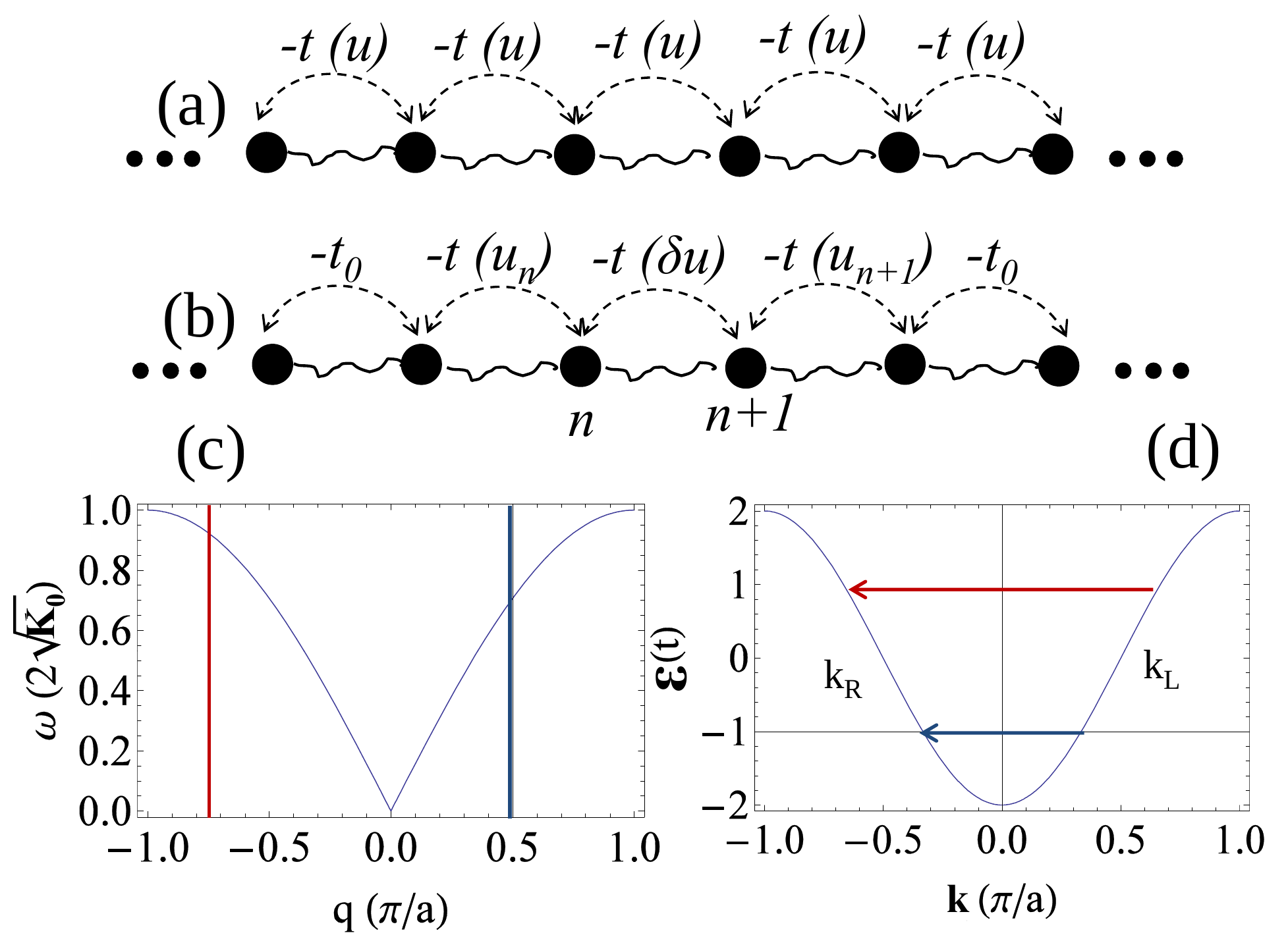}
\caption{(a) A model 1D lattice, where the electron nearest-neighbor hopping
	amplitude $t(u)$ depends on the atomic displacement $u$. (b)Localized
	e-ph interaction at sites $n$ and $n+1$, $\delta u = u_{n+1}-u_n$. (c)
	Phonon dispersion relation. (d) Electron dispersion relation. Blue and
	red lines depict phonon emission processes due to e-ph interaction. An
	electron at state $k_L$ is scattered by phonons to an state $k_R$ and
	emit one phonon, whose wavevector is denoted by the vertical lines in
	(c).  Depending on the value of $k_L$ and $k_R$, the emitted phonon may
	travel to the right (blue line, normal process) or the left (red line, Umklapp
	process). Normally, the electron energy is much larger than the phonon
	energy, so in this figure we have ignored the electron energy change.}
\label{fig:1dmodel}
\label{fig:seebeck}
\end{figure}

For the ease of understanding the general results in Sec.~\ref{sec:loe}, we now study a
simple 1D atomic chain. The electronic Hamiltonian takes the
tight-binding form, with the hopping matrix element $-t$,
\begin{eqnarray}
	H_{\rm e} = -t\sum_{|i-j|=1} (c^\dagger_i c_{j}+{\rm h.c.}).
	\label{eq:lre}
\end{eqnarray}
The electron dispersion relation is $\varepsilon_k = -2t \cos k$, where $k$ is the 1D wavevector[Fig.~\ref{fig:1dmodel} (d)]. We have set the lattice distance $a=1$. 
%The electron DOS is
%\begin{equation}
%	D(\varepsilon) = \frac{1}{\pi}\frac{1}{\sqrt{4t^2-\varepsilon^2}}.
%	\label{eq:edos}
%\end{equation}
For this 1D lattice, due to translational invariance, the electron Green's function in real space only depends on the distance between different sites $j$ and $l$, $n=j-l$,
\begin{equation}
	G_{0,jl}^r(\varepsilon) = \frac{e^{i|k(\varepsilon^+)n|}}{2i t |\sin k(\varepsilon^+)|}.
	\label{eq:gnr}
\end{equation}
Here, $k(\varepsilon^+)$ is the wavevector corresponding to energy $\varepsilon^+=\varepsilon+i 0^+$.
The left and right spectral function, defined within the electron energy band, are
\begin{eqnarray}
	A_{L,jl}(\varepsilon) =\tilde{A}^*_{L,jl}(\varepsilon) = \frac{e^{ik(\varepsilon^+)n}}{2t|\sin k(\varepsilon^+)|},   A_{R}(\varepsilon) = A_{L}^*(\varepsilon).\nonumber\\
	\label{eq:aln}
\end{eqnarray}

The ions are connected by 1D springs  with spring constant $K_0$
\begin{eqnarray}
	H_{\rm p}= \sum_{j}\left(\frac{1}{2}{\dot{u}}_j^2+K_0{u}_j^2\right) -\frac{1}{2}K_0\sum_{|i-j|=1}{u}_i{u}_j,
	\label{eq:lrc}
\end{eqnarray}
The phonon retarded Green's function is
\begin{equation}
	D^r_{0,jl}(\omega) = \frac{e^{i|q(\omega^+)n|}}{2i K_0 |\sin q(\omega^+)|}.
	\label{eq:phdr}
\end{equation}
Here, $q(\omega^+)$ is the phonon wavevector corresponding to frequency $\omega^+=\omega+i 0^+$, which we can get
from the dispersion relation $\omega_q = 2\sqrt{K_0} \left|\sin \frac{q}{2}\right|$ [Fig.~\ref{fig:1dmodel} (c)]. 
The phonon spectral function, defined within the phonon band, is
\begin{equation}
	\mathcal{A}_{L,jl}(\omega) = \mathcal{\tilde{A}}_{L,jl}^*(\omega) = \frac{e^{i q(\omega^+)n}}{2K_0 |\sin q(\omega^+)|}, \mathcal{A}_{R}(\omega) = \mathcal{A}_{L}^*(\omega).
	\label{eq:phls}
\end{equation}

To consider the e-ph interaction, we assume the atomic motion modifies the hopping matrix element linearly, i.e.,  
\begin{equation}
	H_{\rm ep} = -m \sum_{j}{u}_j (c^\dagger_j c_{j+1}-c^\dagger_j c_{j-1}+{\rm h.c.}).
	\label{eq:ephq}
\end{equation}
For a phonon emission process through electronic transition from the initial
left scattering states $|\psi_L(k_L)\rangle$ to the final right scattering
state $|\psi_R(k_R)\rangle$, only phonon mode that fulfills the energy and
crystal-momentum conservation  can be excited, e.g., $\langle
\psi_L(k_L)|M^q|\psi_R(k_R)\rangle \sim \delta(k_L-k_R-q+
G)\delta\bigl(\varepsilon(k_L)-\varepsilon(k_R)-\hbar\omega(q)\bigr)$.
Here, $G$ is a reciprocal lattice vector, and $k_L>0$, $k_R<0$.  It is this selection rule that
gives the electron/phonon drag effect in a translational invariant lattice[Fig.~\ref{fig:1dmodel} (c)-(d)].

To show that similar mechanism works in a coherent
nano-conductor, we artificially switch off the e-ph interaction, except at two
sites, e.g., we only consider coupling to $u_n$ and $u_{n+1}$[Fig.~\ref{fig:1dmodel} (b)].  That is, in Eq.~(\ref{eq:ephq}), the sum over $j$
only applies to these two sites.  Then, only nearest hopping between four sites, 
$\{n-1,n,n+1,n+2\}$, are modified by atomic motion. We set these four sites as our device, and all other sites as electron and phonon
baths $L$ and $R$. In this case, the condition of energy conservation is still
valid, but the conservation of crystal-momentum is not, since the local e-ph interaction breaks the translation invariance. The matrix element is 
\onecolumngrid
\begin{eqnarray}
%	\langle \psi_L(k_L)|M^q|\psi_R(k_R)\rangle&=&-\frac{m}{\hbar\sqrt{|v_L v_R|}}\left( \begin{array}{cccc}1&e^{-ik_L}&e^{-2ik_L}& e^{-3ik_L}\end{array}\right)
%\left( \begin{array}{cccc}
%		0&1&0&0\\
%		1&0&-1+e^{iq}&0\\
%		0&-1+e^{iq}&0&-e^{iq}\\
%		0&0&-e^{iq}&0\\
%	\end{array} \right)
%	\left( \begin{array}{c}
%		1\\
%		e^{ik_R}\\
%		e^{2ik_R}\\
%		e^{3ik_R}
%	\end{array} \right)\nonumber\\ \nonumber\\
	\langle \psi_L(k_L)|M^q|\psi_R(k_R)\rangle &=& -\frac{m }{\hbar\sqrt{|v_L v_R|}}e^{-ik_L}\left[1+e^{i(q-k_L+k_R)}\right]\left[1+e^{i(k_L+k_R)}\right]\left[1-e^{-i(k_L-k_R)}\right].
\end{eqnarray}
\twocolumngrid

Here, $v_{L/R}$ is the group velocity of the $L/R$ scattering state with wavevector $k_{L/R}$.
We can see that the
squared scattering matrix element $|\langle
\psi_L(q_L)|M^q|\psi_R(q_R)\rangle|^2 \neq  |\langle
\psi_L(q_L)|M^{-q}|\psi_R(q_R)\rangle|^2$, their difference
\begin{equation}
	\Delta M_{LR} \propto \sin\phi \sin q.
\end{equation}
We have defined $\phi=k_L-k_R$. This means the electrons have different
probability of exciting left and right traveling phonon waves. The difference
depends on the electron and phonon wavevectors. For example, similar to the 1D
lattice, electrons with $\varepsilon<0$ (below half filling) preferentially
emit phonons travelling to the right. From another point of view, the holes
dominate the inelastic transport\cite{Lu15}.  This breaks the left-right
symmetry, and generates drag effect in a nano-conductor, although the
crystal-momentum selection rule is not valid.

To make connection with the NEGF approach in Sec.~\ref{sec:loe}, we can calculate the real space e-ph interaction matrix at sites $n$ and $n+1$, and find $M^{n+1}_{LR}\equiv\langle \psi_L(k_L)|M^{n+1}|\psi_R(k_R)\rangle=e^{-i\phi}\langle \psi_L(k_L)|M^{n}|\psi_R(k_R)\rangle$.  So,
\begin{eqnarray}
	X^{n,n+1}_{LR}\bigl(\varepsilon(k_L),\varepsilon(k_R)\bigr) =\left|M^n_{LR}\right|^2\left( \begin{array}{cr}
		1 & e^{-i \phi}\\	
		e^{i\phi} & 1 \\	
	\end{array} \right). \label{eq:chi}
\end{eqnarray}
Making use of Eq.~(\ref{eq:phls}), we get, for given electron ($k_L$, $k_R$) and phonon wavevectors ($q$), that satisfy the requirement of energy conservation,
$\varepsilon(k_L)=\varepsilon(k_R)+\hbar\omega(q)$, the electron `drag' term becomes,
\begin{equation}
	{\rm Tr}\bigl[{\rm Im} X_{LR}(\omega) {\rm Im}\tilde{\mathcal{A}}_L(\omega)\bigr] \propto \sin\phi \sin q(\omega^+).
\end{equation}
This is consistent with our scattering analysis, and shows that the drag effect
we discuss here shares the same origin as that in lattice system.

\begin{figure}[h]
\includegraphics[scale=1.0]{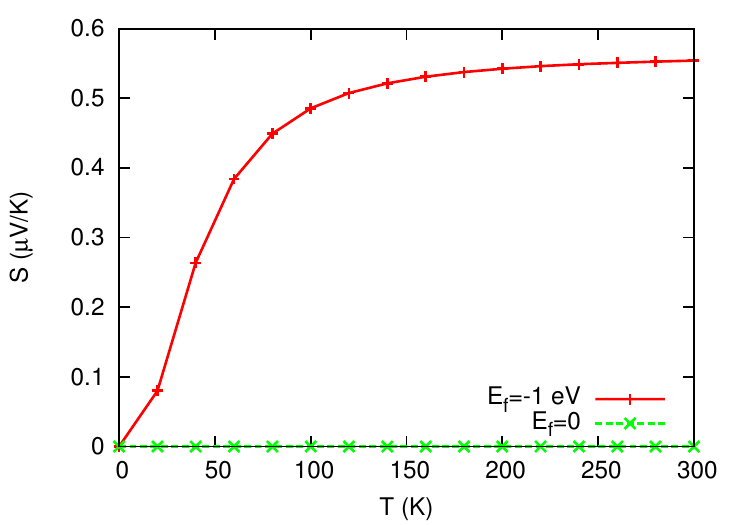}
\caption{Phonon drag contribution to the Seebeck coefficient at different
chemical potentials. The parameters used are as follows: $t=1$ eV, $K_0=0.02$
eV/({\AA}$^2$u), $K'_0=0.5K$, $m=0.15$ eV/(\AA$\sqrt{ {\rm u}}$). Here, $K'_0$
is the spring constant connecting the device to the left and right phonon baths.}
\label{fig:seebeck}
\end{figure}

%%%%%%%%%%%%%%%%%%%%%%%%%%%%%%%%%%%%%%%%%%%%%%%%%%%%%%%%%%%%%%%%%%%%%%%%%%%%%%%%%%
\subsection{Numerical results}
Now we present our numerical results for the 1D model with localized e-ph interaction using the formulas developed in Sec.~\ref{sec:loe}.
In Fig.~\ref{fig:seebeck}, we show the calculated phonon drag contribution to the Seebeck coefficient. The parameters, given in the figure caption, are chosen to closely resemble that of a single atom gold chain\cite{Yeyati2006,Ludoph99,Matsushita2015,Evangeli2015}.  The single
electron contribution to the Seebeck coefficient vanishes, since we have a perfect electron transport
channel.  The drag coefficient is zero at $E_F=0$ due to electron-hole symmetry\cite{Lu15}.  But once moving to $E_F=-1$ eV, the symmetry is broken, and we get a non-zero value. Positive $S$ means the holes dominate  over the electrons in the inelastic scattering process. The saturation of the $S$ with $T$ can be understood from Eq.~(\ref{eq:corQwb}), i.e., at high temperature, $\tilde{\mathcal{Q}}_0 \propto T$, while $S \propto\tilde{\mathcal{Q}}_0/T$.

\begin{figure}[h]
\includegraphics[scale=1.0]{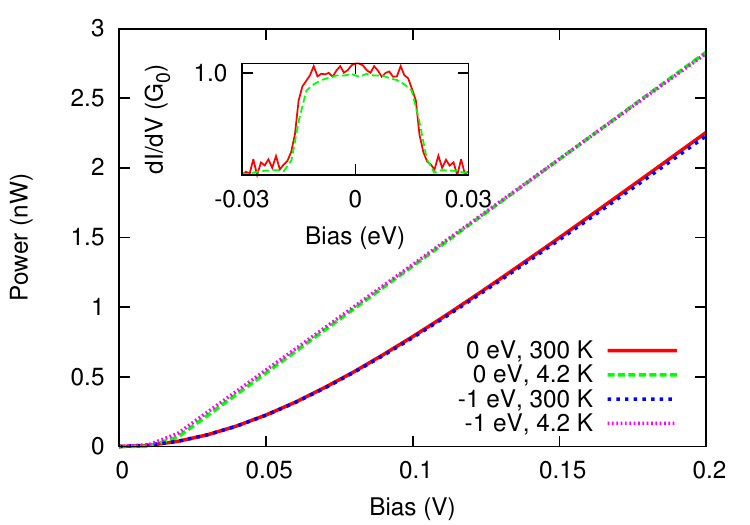}
\caption{Energy current going from electron to phonons (power) as a function of voltage bias at different temperatures and Fermi levels from Eqs.~(\ref{eq:power}-\ref{eq:econs}). All the electron and phonon baths are kept at the same temperature, see also Fig.~\ref{fig:sys} (b). The inset shows the differential conductance ($dI/dV$) at $E_F=0$ (red, solid) and $-1$ eV (green, dashed) at $4.2$ K. }
\label{fig:dIdV}
\end{figure}

\begin{figure}
\includegraphics[scale=1.1]{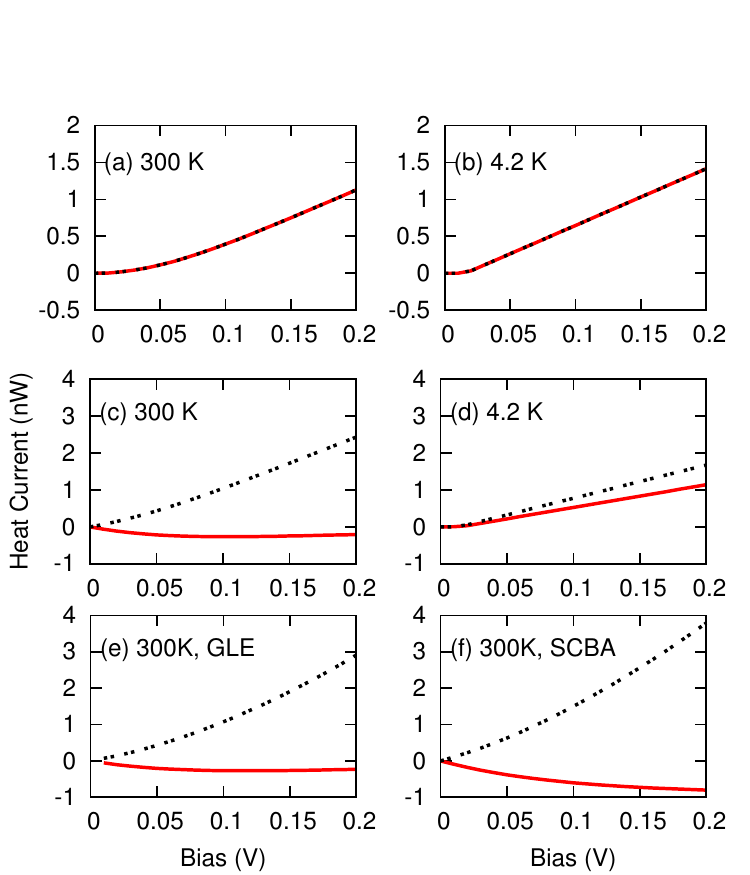}
\caption{Heat current carried by phonons, going into the left (red, solid) and right (black, dashed)
phonon bath. (a)-(b) $E_F=0$, (c)-(f) $E_F=-1$ eV. (a)-(d) Calculated from Eqs.~(\ref{eq:power}-\ref{eq:econs}). We also show the results calculated from the generalized Langevin equation (GLE) approach (Ref.~\onlinecite{luprb12}) (e) and the self-consistent Born approximation (SCBA) (Ref.~\onlinecite{Lu07}) (f). } 
%\caption{Heat current carried by phonons, going into the left and right phonon bath. (a) $E_F=0$, (b) $E_F=-1$ eV. In (b) we also show the results calculated from the self-consistent Born approximation (SCBA) (Ref.~\onlinecite{Lu07}) and the generalized Langevin equation (GLE) approach (Ref.~\onlinecite{luprb12}). We can see, although the magnitude of the heat current changes, the qualitative conclusions remain no matter which approximation we use. \revision{Too many curves.}} 
\label{fig:heat}
\end{figure}

Figures~\ref{fig:dIdV}-\ref{fig:heat} show the results for the setup in
Fig.~\ref{fig:sys} (a). The temperatures of all the baths are the same, while
there is a voltage bias applied between the two electron baths.  We define the power as
Q in Eq.~(\ref{eq:power}). Figure~\ref{fig:dIdV} shows its dependence on the voltage bias at
different temperatures and Fermi levels.  The onset of power flow at the phonon
frequency is clear at $4.2$ K, but smoothed out at $300$ K. The inset shows the
corresponding conductance drop at the phonon threshold for $4.2$ K. Although the
magnitude of the power changes slightly, the general behavior does not depend
on the position of the Fermi level. All these results agree with previous studies\cite{FrBrLoJa.2004,ViCuPaHa05,Lu07}.

Next, we show in Fig.~\ref{fig:heat}(a) and (b) that the heat flows into the
left and right phonon baths are drastically different at the two Fermi levels.
For $E_F=0$, the heat flow into the two phonon baths are symmetric (Fig.~\ref{fig:heat} (a-b)). But when we
move to $E_F=-1.0$ eV, they show strong asymmetry, due to the drag part of the
heat current (Fig.~\ref{fig:heat} (c-d)). It depends on the phase of the electron wavefunction, which in
our case could be tuned by changing the chemical potential. At $E_F=-1$ eV, the
probability of emitting right travelling phonons is larger, resulting in larger
heat current into bath $R$.  This is the same with the lattice system
[Fig.~\ref{fig:1dmodel} (c)-(d)].  Comparing results at $T=4.2$ and $300$ K, we
find that the asymmetry increases with temperature. Interestingly, at
$300$ K, we can extract heat from the right phonon bath by applying the voltage
bias. This is a prototype atomic `refrigerator'. 
The results calculated using the generalized Langevin equation (GLE) approach\cite{luprb12}[Fig.~\ref{fig:heat} (e)] and self-consistent Born approximation (SCBA)\cite{Lu07}[Fig.~\ref{fig:heat} (f)] show that, although the magnitude of the heat current changes, the qualitative conclusions remain no matter which approximation we use.

\section{Conclusions}
\label{sec:conc}
In conclusion, by assuming linear coupling between electron and phonons in a
four-terminal nano-device, we have shown that in the linear transport regime,
in addition to modifying the normal thermoelectric transport coefficients, e-ph
interaction also introduces new drag type coefficients. The drag effect can be traced back to the momentum transfer between the
electrons and phonons. We have shown that it is closely related to the
adiabatic pumping, and current-induced forces in a coherent conductor. So
in principle phonon-drag thermopower behaves as an alternative way of probing
these current-induced forces. The expressions derived in this paper
can be readily applied to the realistic structures by combining these with
first-principles electronic structure calculation\cite{Soler.02,BrMoOr.2002,FrPaBr.2007}.
Finally, we note that one could also study similar drag effect in coulomb
coupled all-electronic devices\cite{Sanchez2010,Sothmann2012,Whitney2016}.

\begin{acknowledgments}
JTL is supported by the National Natural Science Foundation of China (Grant
Nos. 11304107 and 61371015).  J.-S. W acknowledges support of an FRC grant
R-144-000-343-112. M. B. acknowledges support from Center for
Nano-structured Graphene (Project DNRF58).
\end{acknowledgments}

\onecolumngrid
\appendix

\section{Expressions for $\mathcal{L}$ and $\mathcal{K}_{\rm p}$}
\label{sec:ephcof}
We write down the full expressions for the thermoelectric coefficients in Eq.~(\ref{eq:lang2}).
The complete $\mathcal{L}$ including e-ph interaction is made from three contributions:
\begin{eqnarray}
	\mathcal{L}_n=\sum_{i=1}^3\mathcal{L}^{(i)}_n,
	\label{eq:cor}
\end{eqnarray}
with
\begin{eqnarray}
	\mathcal{L}^{(1)}_n &=& \frac{1}{\hbar}\int \frac{d\varepsilon}{2\pi} (\varepsilon-\mu)^n \textrm{tr}\left[A_{\bar{\alpha}}(\varepsilon)\Gamma_\alpha(\varepsilon)\right]f'(\varepsilon),\quad
	\mathcal{L}^{(2)}_n = \frac{1}{\hbar}\int \frac{d\varepsilon}{2\pi} (\varepsilon-\mu)^n \textrm{tr}\left[\Delta A_{\bar{\alpha}}(\varepsilon)\Gamma_\alpha(\varepsilon)\right]f'(\varepsilon),\\
	\label{eq:cor23}
	\mathcal{L}^{(3)}_n &=& \frac{i}{\hbar}\int \frac{d\varepsilon}{2\pi}(\varepsilon-\mu)^n {\rm tr}\bigl[(\Sigma_{\rm ep}^>(\varepsilon)-\Sigma_{\rm ep}^<(\varepsilon))\tilde{A}_\alpha(\varepsilon)\bigr]f'(\varepsilon).
\end{eqnarray}
We have defined $f'(\varepsilon)=-\frac{\partial
f(\varepsilon,\mu,T)}{\partial\varepsilon}$.  Since the chemical potential
$\mu$ and temperature $T$ are all the same for both electrodes. We have omitted 
them to simplify the expressions. We also have
\begin{eqnarray}
	\label{eq:ea}A_\alpha(\varepsilon)=G_0^r(\varepsilon)\Gamma^e_\alpha(\varepsilon) G_0^a(\varepsilon),\quad
	\label{eq:ear}\tilde{A}_\alpha(\varepsilon)=G_0^a(\varepsilon)\Gamma^e_\alpha(\varepsilon) G_0^r(\varepsilon),\quad
	A(\omega)=A_L(\omega)+A_R(\omega), \\
	\label{eq:pha}{\mathcal{A}}_\alpha(\omega)=D_0^r(\omega)\Gamma^{\rm p}_\alpha(\omega) D_0^a(\omega),\quad 
	\tilde{\mathcal{A}}_\alpha(\omega)=D_0^a(\omega)\Gamma^{\rm p}_\alpha(\omega) D_0^r(\omega), \quad \label{eq:a}\mathcal{A}(\omega)=\mathcal{A}_L(\omega)+\mathcal{A}_R(\omega).
\end{eqnarray}
$\mathcal{L}^{(1)}_n$ is the single electron Landauer result.
$\mathcal{L}^{(2)}_n$  is due to corrections to the
electron DOS
\begin{eqnarray}
	\Delta A_{\bar{\alpha}} = G^r_0(\varepsilon)\Sigma_{\rm ep}^r(\varepsilon)A_{\bar{\alpha}}(\varepsilon)+A_{\bar{\alpha}}(\varepsilon)\Sigma_{\rm ep}^a(\varepsilon)G_0^a(\varepsilon).
\end{eqnarray}
$\mathcal{L}^{(3)}$ is the inelastic term. 
Note that only the Fock diagram contributes to $\mathcal{L}^{(3)}$.
Effect of e-ph interaction on the $\mathcal{L}_n$ in a single level model has been studied 
in Ref.~\onlinecite{Lu2015Rev}.

The phonon thermal conductance has similar form
\begin{equation}
	\mathcal{K}_{\rm p}=\sum_{i=1}^3 \mathcal{K}^{(i)}_{\rm p}
	\label{eq:phth}
\end{equation}
with
\begin{eqnarray}
	\mathcal{K}^{(1)}_{\rm p}&=&\int \frac{d\omega}{4\pi}\hbar\omega \textrm{Tr}\left[\mathcal{A}_{\bar{\alpha}}(\omega)\Gamma_{\rm p}^{\alpha}(\omega)\right]\frac{\partial n_B}{\partial T_{\rm p}}T_{\rm p}, \label{eq:kcor1}\quad
	\mathcal{K}^{(2)}_{\rm p}=\int \frac{d\omega}{4\pi}\hbar\omega \textrm{Tr}\left[\Delta\mathcal{A}_{\bar{\alpha}}(\omega)\Gamma_{\rm p}^{\alpha}(\omega)\right]\frac{\partial n_B}{\partial T_{\rm p}}T_{\rm p},\\
	\mathcal{K}^{(3)}_{\rm p}&=&i\int \frac{d\omega}{4\pi}\hbar\omega \textrm{Tr}\left[(\Pi_{\rm ep}^>(\omega)-\Pi_{\rm ep}^<(\omega))\tilde{\mathcal{A}}_\alpha(\omega)\right]\frac{\partial n_B}{\partial T_{\rm p}}T_{\rm p}.
	\label{eq:kcor234}
\end{eqnarray}
Here $\bar{\alpha}$ means electrode opposite to $\alpha$.
Similarly, $\mathcal{K}^{(2)}_n$ is due to corrections to the phonon DOS
\begin{equation}
	\Delta \mathcal{A}_{\bar{\alpha}} = D^r_0(\varepsilon)\Pi_{\rm ep}^r(\varepsilon)\mathcal{A}_{\bar{\alpha}}(\varepsilon)+\mathcal{A}_{\bar{\alpha}}(\varepsilon)\Pi_{\rm ep}^a(\varepsilon)D_0^a(\varepsilon).
\end{equation}

\section{Onsager symmetry}
\label{sec:ons}
Here, we show that $\mathcal{Q}_n = \tilde{\mathcal{Q}}_n$  in the presence of
time-reversal symmetry, e.g., $(G^r)^T = G^r$. In that case, from Eq.~(\ref{eq:pha}), we have
\begin{equation}
  \tilde{\mathcal{A}}_\alpha = \mathcal{A}_\alpha^*,\quad \tilde{\mathcal{A}} = \mathcal{A},\quad \tilde{{A}}_\alpha = {A}_\alpha^*, \quad \tilde{A} = A.
  \label{}
\end{equation}
Furthermore, $A$ and $\mathcal{A}$ are both real.
Consequently, in the linear response regime, we can set $f(\varepsilon,\mu_\alpha,T_\alpha) = f(\varepsilon,\mu_\beta,T_\beta)$ in Eq.~(\ref{eq:lamab}), and get
\begin{equation}
  \sum_\beta \Lambda_{\tilde{\alpha}\beta}^{(n)} =  \sum_\beta \Lambda_{\tilde{\alpha}\tilde{\beta}}^{(n)} = \big(\sum_\beta \Lambda_{{\alpha}{\beta}}^{(n)}\big)^*.
  \label{}
\end{equation}
Using the above two equations, we see that $\mathcal{Q}_n^* =
\tilde{\mathcal{Q}}_n$. 

On the other hand, the coefficients $\mathcal{Q}_n$ are
real $\mathcal{Q}_n=\mathcal{Q}_n^*$. This can be seen from the fact that: 
(1) the matrices $M$, $A$, $\mathcal{A}$ are real, and $A_\alpha/\mathcal{A}_\alpha$ is Hermitian; (2) the trace of their products is real. Putting them together, we get the desired result $\mathcal{Q}_n = \tilde{\mathcal{Q}}_n$.

\section{Equivalence to the results of Ref.~38}
\label{sec:eew}
Here, we show that  results of
Ref.~\onlinecite{entin-wohlman_three-terminal_2010} (Eqs.~(35-36)) are special
case of our results. There, the authors considered one electronic level at
($\varepsilon_0$) coupled to one vibrational mode with angular frequency
$\omega_0$. The electronic level couples to two electrodes and the vibrational
mode couples to one phonon bath, characterized by $\gamma_{ph}$. In this
special case, all the matrices become numbers. We get
$G_0^r(\varepsilon) = \left[\varepsilon+i \gamma(\varepsilon)/2-\varepsilon_0\right]^{-1}$, with $\gamma(\varepsilon)=\gamma_R(\varepsilon)+\gamma_L(\varepsilon)$,
and $D_0^r(\omega) = \left[(\omega + i \gamma_{\rm ph})^2 - \omega_0^2\right]^{-1}$.
Now we have $\mathcal{Q}_0=0$, since the phonon mode couples only to one bath and $\mathcal{A}_\alpha$ is real.
Substituting these two equations to Eq.~(\ref{eq:corQ}), and assuming
$\gamma_{\rm ph}$ is small, we get
\begin{eqnarray}
  \label{eq:cew}
  \mathcal{Q}_n &=& m^2\int \frac{d\varepsilon}{4\pi} (\varepsilon+{\hbar\omega_0}/{2}-\mu)^n {|G^r_0(\varepsilon+\hbar\omega_0/2)|^2|G^r_0(\varepsilon-\hbar\omega_0/2)|^2}\\
  &\times&\partial_{\hbar\omega}n_B(\hbar\omega_0) \left[\gamma_\alpha(\varepsilon-\hbar\omega_0/2)\gamma_{\bar{\alpha}}(\varepsilon+\hbar\omega_0/2)- \gamma_\alpha(\varepsilon+\hbar\omega_0/2)\gamma_{\bar{\alpha}}(\varepsilon-\hbar\omega_0/2) \right](f(\varepsilon+\hbar\omega_0/2)-f(\varepsilon-\hbar\omega_0/2))  \nonumber.
\end{eqnarray}
This is consistent with Eqs.~(35-36) of
Ref.~\onlinecite{{entin-wohlman_three-terminal_2010}}. The extra factor $1/2$
in Eq.~(\ref{eq:cew}) comes from different definition of the e-ph
interaction $m$. Note that we need $\gamma_\alpha(\varepsilon)$ to be energy dependent, and
$\gamma_\alpha(\varepsilon)\neq\gamma_{\bar{\alpha}}(\varepsilon)$, in order
for $\mathcal{Q}_n \neq 0$.
\twocolumngrid
%\bibliography{thermoelectric}
%
\end{document}